\documentclass[a4paper]{article}

\usepackage[english]{babel}
\usepackage[utf8x]{inputenc}
\usepackage[T1]{fontenc}
\usepackage{authblk}

\usepackage[a4paper,top=3cm,bottom=2cm,left=3cm,right=3cm,marginparwidth=1.75cm]{geometry}

\usepackage{amsmath}
\usepackage{graphicx}
\usepackage[colorinlistoftodos]{todonotes}
\usepackage[colorlinks=true, allcolors=blue]{hyperref}
\usepackage{subfig}

\title{Differential Imaging Forensics}
\author[1,2]{Aur\'elien Bourquard}
\author[3]{Jeff Yan}
\affil[1]{Research Laboratory of Electronics, Massachusetts Institute of Technology, USA} 
\affil[2]{Biomedical Image Technologies, Universidad Polit\'ecnica de Madrid and CIBER-BBN, Madrid, Spain}
\affil[3]{IDA, Link\"opings Universitet, Sweden}
\affil[ ]{\href{mailto:aurelien@mit.edu}{aurelien@mit.edu}; \href{mailto:jeff.yan@liu.se}{jeff.yan@liu.se}}

\begin{document}
\maketitle

\begin{abstract}

We introduce some new forensics based on differential imaging, where a novel category of visual evidence created via subtle interactions of light with a scene, such as dim reflections, can be computationally extracted and amplified from an image of interest through a comparative analysis with an additional reference baseline image acquired under similar conditions. This paradigm of \emph{differential imaging forensics} (DIF) enables forensic examiners for the first time to retrieve the said visual evidence that is readily available in an image or video footage but would otherwise remain faint or even invisible to a human observer. 
We demonstrate the relevance and effectiveness of our approach through practical experiments. We also show that DIF provides a novel method for detecting forged images and video clips, including deep fakes. 

\end{abstract}

\section{Introduction}
\label{Section: Introduction}

Given a {\it single} photo, how to determine who was the cameraman? This was a question we first raised in \cite{yan2017poster}, motivated by its relevance to intelligence communities, law enforcement and privacy research. For example, assume that an insider took a photo of a secret military facility in Russia and gave it to the Central Intelligence Agency. Once the photo got leaked by a mole inside the CIA, Russian anti-spy operatives would be keen to figure out who took the photo in the first place. Similarly, if the law enforcement was tipped off for a crime by a photo from an anonymous source, they might want to identify the person behind the camera to gain further clues. The question that we asked is, in general, hard to resolve, except for several special cases.

A related question that we ask is the following.
For convenience of discussions but without loss of generality, consider the investigation of a case where a high-value merchandise has been stolen from an indoor environment\footnote{Concepts and methods introduced in this paper, together with all discussions to follow up, are applicable to outdoor settings, too}. In this case, no material evidence is found in the scene, but the police has apprehended a suspect who is known to regularly wear a unique type of green jacket. If this garment were identified from an image of one of the surveillance cameras, this would help incriminate the suspect and solve the case. Unbeknownst to the investigators, however, the thief had managed to identify the position of all surveillance cameras and stay out of their field of view (FoV), thus compromising such a direct identification. Can we offer innovative techniques to help the law enforcement investigate such cases? 
%
%
%


In this paper, we address the above questions, and propose a method that extracts and identifies visual evidence that was created via subtle interactions of light with the scene, which no prior art has done. Our contributions are:

\begin{itemize}

\item The uncovering of a new category of visual evidence in images and video footage, which no current forensic methods can provide. 
%
%

\item The concept of \emph{differential imaging forensics} (DIF), which sets a new paradigm for extracting the said novel category of visual evidence from recorded visual data. This concept involves the use of an additional \emph{reference baseline image} that captures the same scene under the same acquisition conditions, but in a controlled setting where it is known that no hidden individuals or objects are present.

\item A new computational technique that extracts and amplifies this new category of forensic evidence from an image of interest so as to be of use to investigators, based on a comparative analysis of this image and the reference baseline image. This technique works on 
both single images and videos, and is designed in such a way that the visual evidence, otherwise hidden and invisible, can be extracted by following a simple, practical and repeatable procedure.

\item A new way to verify whether an image or video clip has been tampered with, based on the proposed DIF methodology.
\end{itemize}

Building on prior concepts and simulation studies we reported \cite{yan2017poster}, this work is the first to introduce the notion and coin the term of \emph{differential imaging forensics}, generalize its application scenarios, and validate them through practical experiments. 

In contrast to DIF, a form of differential analysis has been widely practiced in digital forensics, by examining digital artefacts (such as computer files, disk images, memory dumps, and network packet dumps) in a pairwise fashion  \cite{garfinkel2012general}. This 
practice allows forensic investigators to focus on what need be paid attention to, but it does not identify or extract any novel genre of evidence.

%

In Section \ref{Section: ProposedApproach}, we describe our approach to recovering 
novel forensic evidence from an 
image. 
In particular, we introduce the concept of a \emph{reference baseline image} that reproduces the acquisition settings of an input scene image of interest, and describe our method of differential analysis. In Section \ref{Section: Experiments}, we conduct practical experiments where our approach is evaluated in different settings. In Section \ref{Section: VideoApplications}, we generalize our approach to the analysis of video footage. In Section \ref{Section: ForgeryDetection}, we discuss why the DIF concept constitutes a novel way to detect forged images and video. In Section \ref{Section: Discussion}, we discuss the implications of our experimental results, further applications of our work, and future research directions. We conclude the paper in Section \ref{Sect: Conclusions} by putting DIF in perspective.

\section{Methodology}
\label{Section: ProposedApproach}

As discussed in Section \ref{Section: Introduction}, images may contain visual evidence that is not directly identifiable by forensic investigators. The methodology we propose aims at retrieving as much of this evidence as possible, when present, to help the investigators identify objects or individuals of interest. It focuses on scenarios where:

\begin{itemize}
\item One or several images of a scene are available for forensic analysis by the investigators
\item Said images have been acquired at a time when potential individuals or objects of interest could be present near the scene
\item Said images contain no---or insufficient---visual evidence that can be located and identified as such by the investigators in the FoV
\end{itemize}
Below, we shall focus on the case of a single scene image, knowing that the proposed methodology is also applicable to cases involving multiple scene images or videos, as shown in Section \ref{Section: VideoApplications}.

\subsection{Approach: a high-level view}

Any object or individual can leave a trace in the digital image of a scene by virtue of its impact on the surrounding light distribution, even if it is located outside of the FOV of that image. In computer graphics, this principle is accounted for in the rendering equation, any two locations in the scene being potentially interdependent \cite{kajiya1986rendering}. When the effect of an object on a scene is either a dark shadow or a mirror-like reflection, it is visible upon simple inspection.

However, more subtle effects may only be uncovered through more advanced methods. Accordingly, we shall specifically focus on effects whose magnitude is too low to be perceived by human observers yet large enough to leave some traces in a digital image. By definition, such effects will cause specific color and brightness changes in the image that are too small to be perceived, isolated, or identified by a human observer. This raises the following technical question: \textbf{If the presence of an object or individual in a scene causes subtle changes in a given scene image, how can one extract and amplify these changes so that they can be visualized?}

This category of latent visual evidence has never been studied before, and we propose the following approach to recover them for the first time.

\begin{itemize}
\item \textbf{Introducing a reference baseline image $p_r$:} This image must accurately reproduce the already-available scene image $p$, except for the subtle color and brightness changes that are characteristic to individuals or objects of interest that were present in the scene when the image $p$ was acquired. Accordingly, the acquisition of the baseline image must be done with a camera of the same type and configuration including the distance between the camera and the scene, point of view, lighting conditions, etc., but with the potential individuals or objects to be identified absent from the scene. The reference baseline image may be acquired \emph{post-hoc} by the investigators, or it may readily be available in some cases, as discussed in Section \ref{Section: VideoApplications}.

\item \textbf{Differential image analysis:} Based on a numerical method detailed in Section \ref{subsect: DiffImageAnalysis}, the differences between the scene image $p$ and the corresponding reference baseline image $p_r$, even if imperceptible, can be extracted and amplified computationally. The amplification process that follows reveals these differences by making them perceptible to human observers.

\item \textbf{Inspecting extracted evidence:} The traces, once extracted and amplified, might offer visual evidence which the investigators inspect to find cues on potential objects or individuals of interest that were present in the scene. This may lead to anything between (a) merely inferring whether one or several objects or individuals were present in the scene or not -- essentially a yes-or-no inference, and (b) identifying one or several individuals or objects that were present in the scene -- this outcome is likely if a sufficient amount of characteristics can be recognized.
\end{itemize}

\subsection{Differential Image Analysis Method}
\label{subsect: DiffImageAnalysis}

Here, we introduce the computational analysis method that we have designed to extract and amplify latent visual evidence from a scene image $p$ with the help of the corresponding reference baseline image $p_r$. Our method essentially includes first extracting subtle brightness or color differences that exist between $p$ and $p_r$, and then enhancing these differences through spatial filtering.

Without loss of generality, we assume that $p$ and $p_r$ are digital images of same resolution $W \times H$, where $W$ and $H$ are the width and the height in pixels, respectively. Introducing our notation, a digital image $i$ maps every color channel $c$ of every pixel $\mathbf{k} = (k_1, k_2)$ with a specific brightness value $i[c, \mathbf{k}]$, where $k_1 \in \{1 \ldots W\}$ is the horizontal coordinate, and where $k_2 \in \{1 \ldots H\}$ is the vertical coordinate.

Based on both images, our differential-analysis method first creates the \emph{difference image} $d$ that is defined for every channel $c$ and pixel $\mathbf{k}$ by the channel-wise and pixel-wise subtraction:
\begin{equation}
\label{eq: Differentiation}
d[c, \mathbf{k}] = p[c, \mathbf{k}] - p_r[c, \mathbf{k}].
\end{equation}
By construction, the difference image $d$ isolates the visual evidence that is associated with elements that are present in the scene at the time the image of interest $p$ was acquired but that are not present in the reference baseline image $p_r$, which is consistent with our strategy discussed above.

While $d$ contains all the differential information of interest, its contrast is typically low, compared to the level of camera noise. Thus, $d$ may be hard to analyze for a human observer, even if its values were amplified. To address this issue, we further enhance the contrast-to-noise ratio of $d$ by applying a spatial-filtering process. The underlying motivation to do so is that salient colors that are associated with objects or individuals, unlike noise, tend to correlate spatially over significant image areas. Accordingly, we define
\begin{equation}
\label{eq: Integration}
D = d \star G_{\sigma}
\end{equation}
where $G_{\sigma}$ is a two-dimensional Gaussian filter of standard deviation $\sigma$ in pixels, and $\star$ denotes the discrete convolution operator acting separately on each channel of $d$. Note that this enhancement process is not needed if the contrast-to-noise ratio is sufficient. In previous simulation studies \cite{yan2017poster}, for instance, only pixel-wise subtraction was used. Finally, based on $D$, we create two difference images $D_{+}$ and $D_{-}$ that amplify the dynamic range of $D$ and are restricted to the positive and negative brightness differences, respectively. These images are defined as follows.
\begin{eqnarray}
\label{eq: ContrastNorm}
\nonumber D_{+}[c, \mathbf{k}] &=& A_{+} \max(D[c, \mathbf{k}], 0),\\
D_{-}[c, \mathbf{k}] &=& A_{-} \min(D[c, \mathbf{k}], 0),
\end{eqnarray}
where $A_{+}$ and $A_{-}$ are scalars normalizing contrast such that the maximum positive and negative brightness differences in $D_{+}[c, \mathbf{k}]$ and $D_{-}[c, \mathbf{k}]$ equal to $1$, which corresponds to, as we assume by convention, the highest possible brightness level on a display. In both resulting images, values are thus restricted to the range $[0, 1]$, with $0$ denoting no difference between the images $p$ and $p_r$. The rationale for separately analyzing positive and negative brightness differences is that they tend to be associated with distinct effects. Since light energy is a positive quantity, the brightness differences in $D_{+}$ are associated with additional light impacting the scene in specific areas, which may correspond to reflections of objects or individuals that were present when $p$ was acquired. Conversely, the brightness differences in $D_{-}$ are associated with decreases in light intensity, which may correspond to shadows or occlusions.

\section{Still Image Experiments}
\label{Section: Experiments}

In this section, we report empirical experiments on single images involving real scenes and objects. While an exhaustive coverage of all types of case scenarios lies beyond the scope of this paper, we address two instances where visual evidence that is too faint to be located and identified by human observers is to be retrieved with our DIF method. 

The scene images were acquired with an iPhone 6 camera of model A1586 through the standard \emph{Camera} application. In both experiments, both acquisitions $p$ and $p_r$ were done with a tripod from the same location and with the same camera settings, with both white balance and focus locked. The F number was 2.2 and the exposure time was $1/17$ second as per standard iPhone settings. The image resolution was $2448 \times 3264$ pixels in RGB format (sRGB IEC61966-2.1). Our analysis method was parameterised with $\sigma = 9$ in Equation (\ref{eq: Integration}).

\subsection{Experiment 1: Uncovering Unexpected Object Presence}
\label{subsect: Exp1}

In this experiment, we address a scenario where an image $p$ of a scene is available, and where we must establish if an unexpected object was present in the scene at the exact time when $p$ was acquired. Based on our strategy, we propose to determine the presence of such an object based on $p$ itself and on an additional reference baseline image $p_r$ that is acquired \emph{post-hoc} in this case.

\begin{figure}
  \centering
  \subfloat[]{\includegraphics[width=0.7\textwidth]{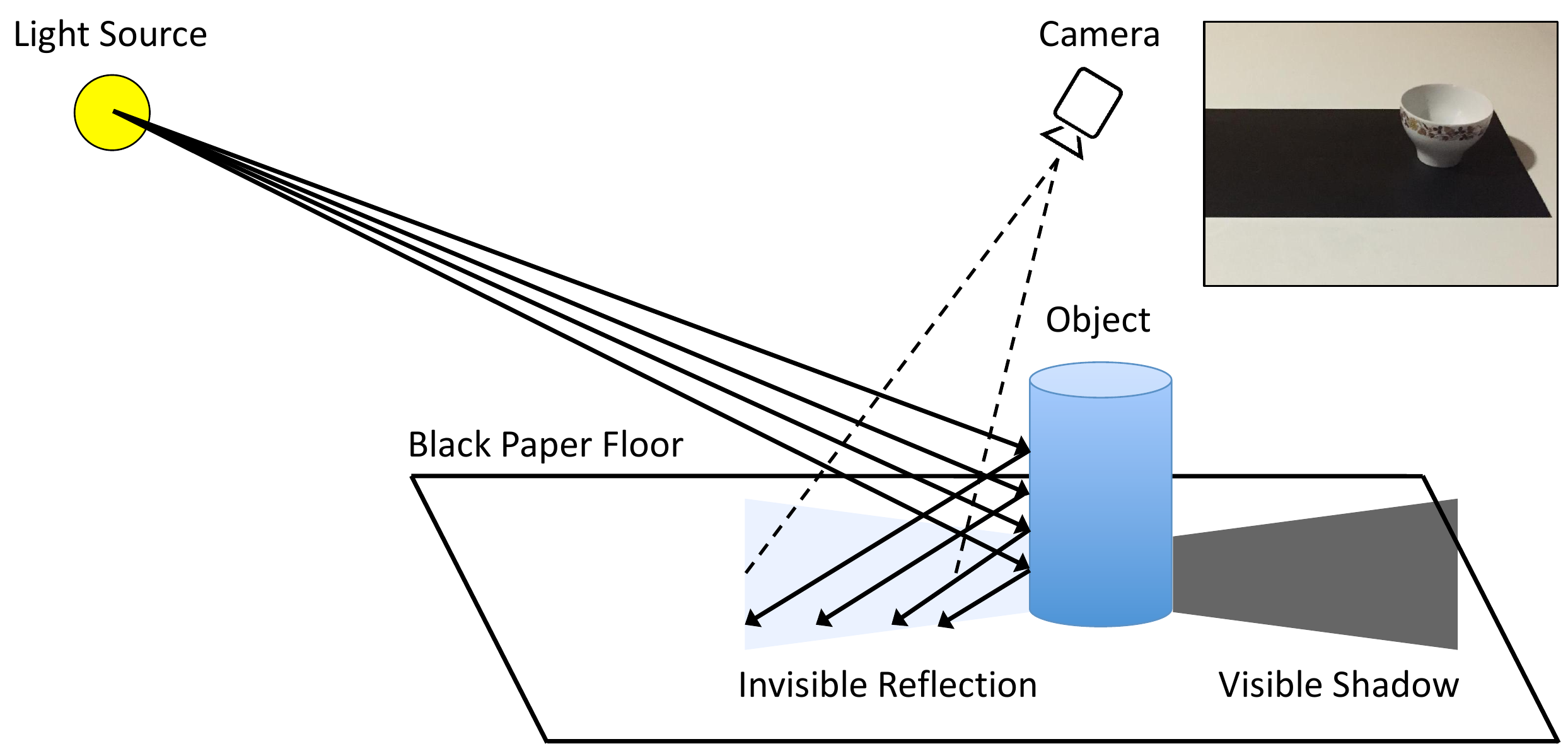}} \hspace{0.1cm}
  \subfloat[]{\includegraphics[trim={0 0 0 60cm},clip,width=0.4\textwidth]{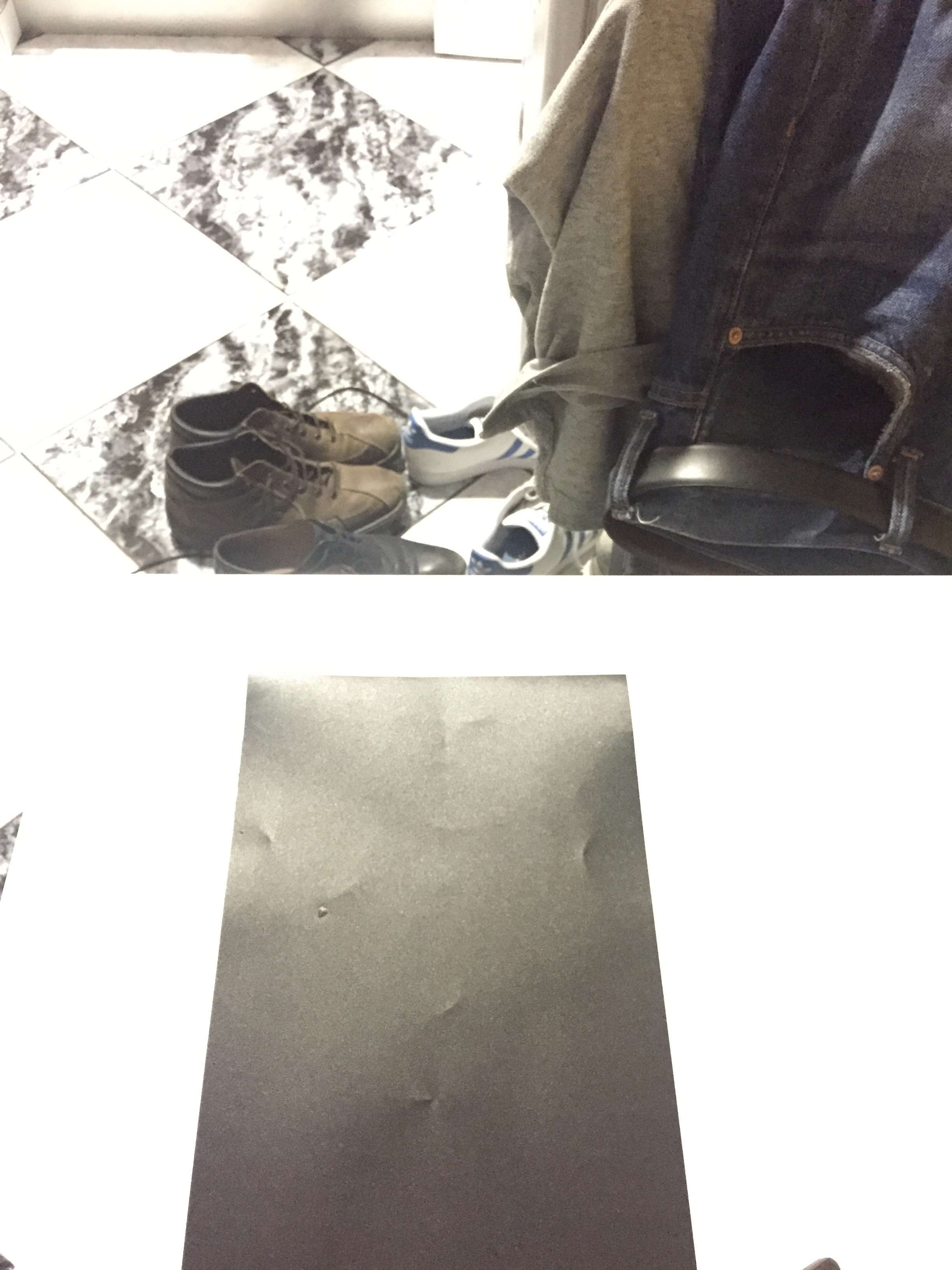}} \hspace{0.1cm}
  \subfloat[]{\includegraphics[trim={0 0 0 60cm},clip,width=0.4\textwidth]{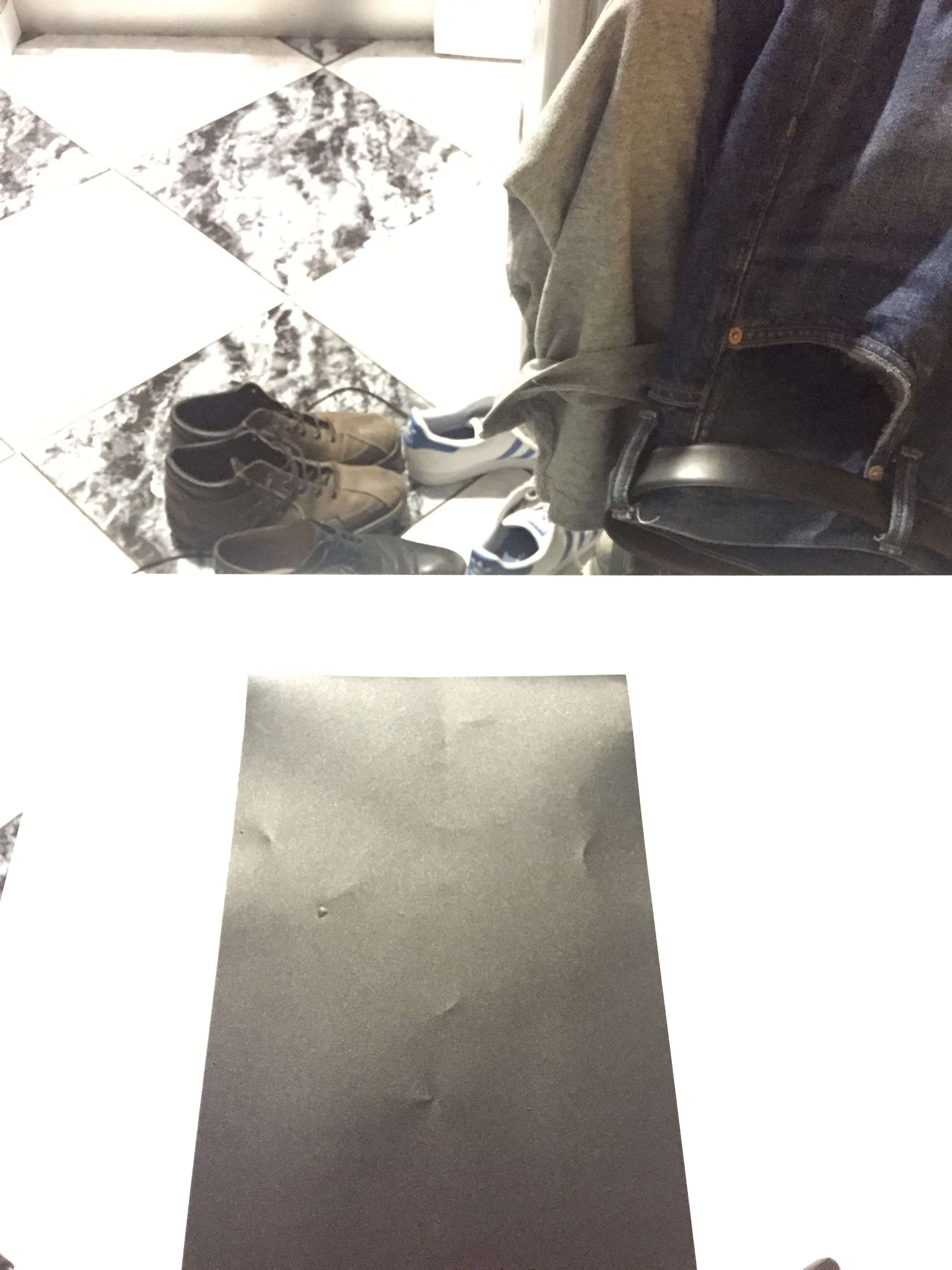}}\\ \vspace{-0.3cm}
  \subfloat[]{\includegraphics[trim={0 0 0 12cm},clip,width=0.4\textwidth]{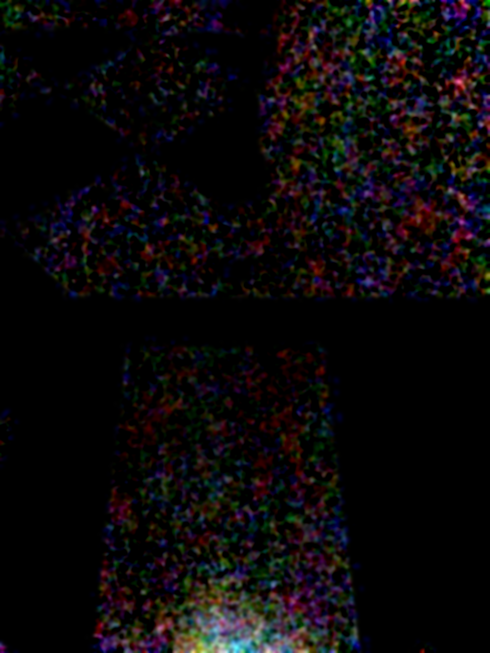}} \hspace{0.1cm}
  \subfloat[]{\includegraphics[trim={0 0 0 12cm},clip,width=0.4\textwidth]{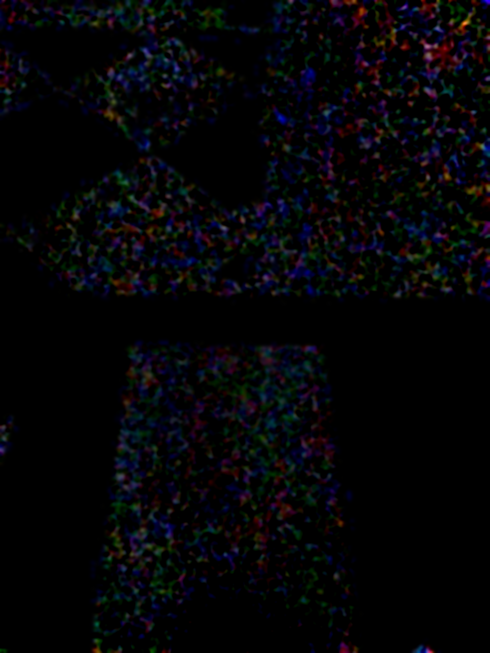}}
  \caption{Uncovering unexpected presence of objects. (a) Side view of the scene, with both schematics and object picture (see inset) shown. The scene includes a dark paper floor, and a cup that was present during the acquisition of $p$ but absent during the acquisition of $p_r$. (b) Available scene image $p$. (c) Reference baseline image $p_r$ acquired post hoc. (d) Difference image $D_{+}$. (e) Difference image $D_{-}$. Difference images in (d) and (e) were cropped out for better visualization. Images and contents thereof are best seen magnified in an e-copy of our paper.}
\label{fig: ResultsCaseStudy2}
\end{figure}

The scene (Figure \ref{fig: ResultsCaseStudy2}a) consists in a dark paper floor, with a cup placed on it during the acquisition of $p$ and removed afterwards. This cup is illuminated by a light source and casts a visible shadow thereof on the floor. However, both the cup and the shadow are located outside the FOV of the camera, thus leaving no visible evidence in $p$ (Figure \ref{fig: ResultsCaseStudy2}b), even upon close inspection. Therefore, only our strategy may reveal additional visual evidence from $p$, potentially revealing the unexpected presence of an object.

Following our strategy, an additional baseline reference image $p_r$ was first acquired in the same conditions as $p$, except for the fact that no object was present in the scene this time. This reference image (\ref{fig: ResultsCaseStudy2}c) displays no visible differences compared to $p$. Subsequently, the corresponding difference images $D_{+}, D_{-}$ were generated following the proposed analysis method described in Section \ref{subsect: DiffImageAnalysis}. Our results show that the positive difference image $D_{+}$ (Figure \ref{fig: ResultsCaseStudy2}d) uncovers the presence of an additional brightness profile at the bottom of the image. In accordance with the positions and orientations of the camera and the light source, this profile was created consequentially as the light emitting from the light source was reflected by the external surface of the cup and then the reflection had an impact on the dark paper floor (\ref{fig: ResultsCaseStudy2}a). By contrast, the negative difference image $D_{-}$ (Figure \ref{fig: ResultsCaseStudy2}e) does not reveal particular details, and it essentially contains leftover camera noise. The fact that no occlusions or similar effects are observed in $D_{-}$ is expected because the visible shadow 
cast by the cup was opposite to the camera orientation.

As shown in this experiment, 
DIF recovers the latent 
evidence that is available in a picture but invisible to human eyes, 
thus revealing the unexpected presence of objects 
located near a scene.

\subsection{Experiment 2: Identifying a Cameraman}
\label{subsect: SecondCaseStudy}

In this experiment, we wish to determine if the proposed strategy not only allows to establish the presence of an unexpected object in a scene as in Section \ref{subsect: Exp1}, but also to clearly identify an object that can lead to the identification of a cameraman. Specifically, inspired by the problem described in Section \ref{Section: Introduction}, our goal is to infer the presence of a green jacket in a scene from a picture $p$ that is made available to investigators. 

\begin{figure}
  \centering
  \subfloat[]{\includegraphics[width=0.227\textwidth]{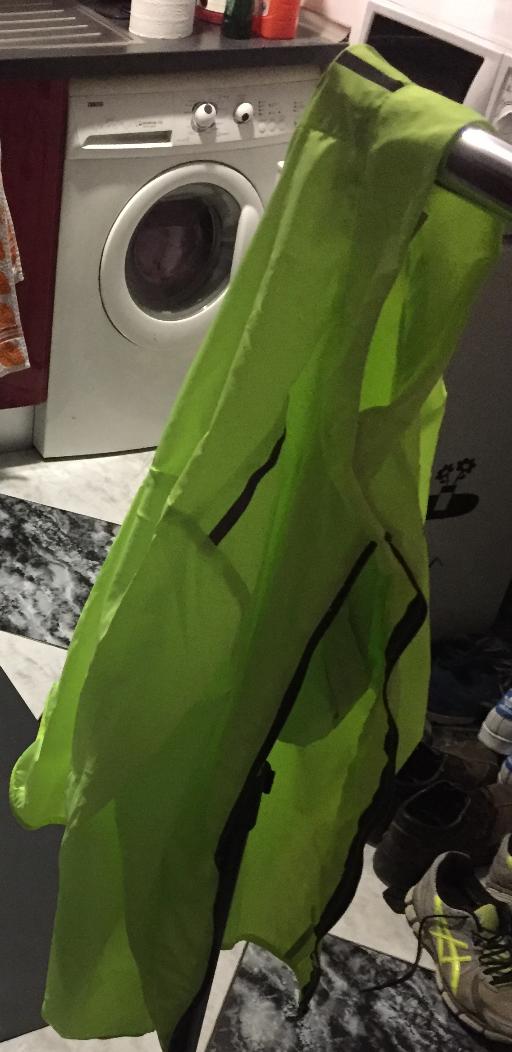}} \hspace{0.1cm}   
  \subfloat[]{\includegraphics[width=0.35\textwidth]{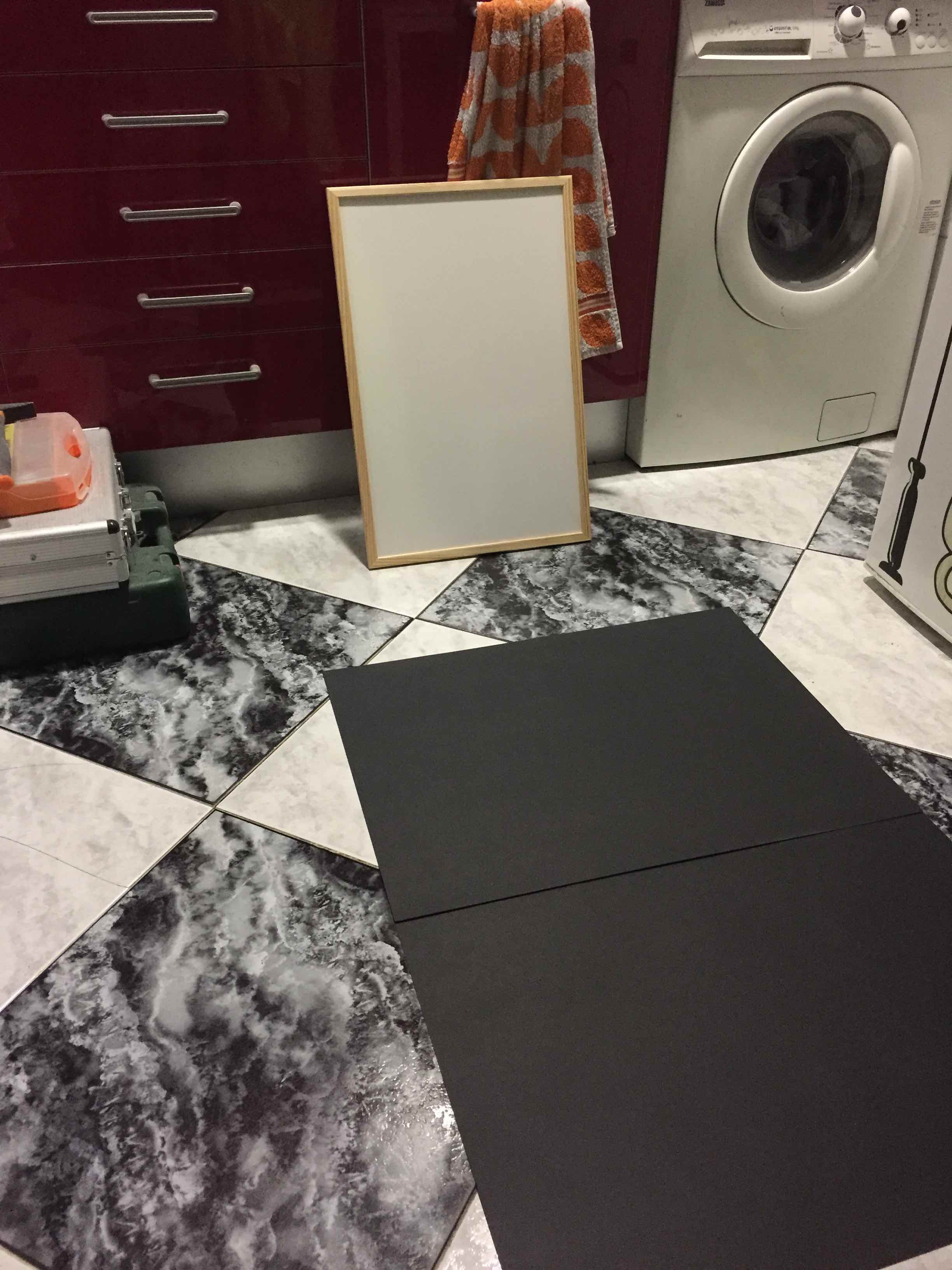}} \hspace{0.1cm}
  \subfloat[]{\includegraphics[width=0.35\textwidth]{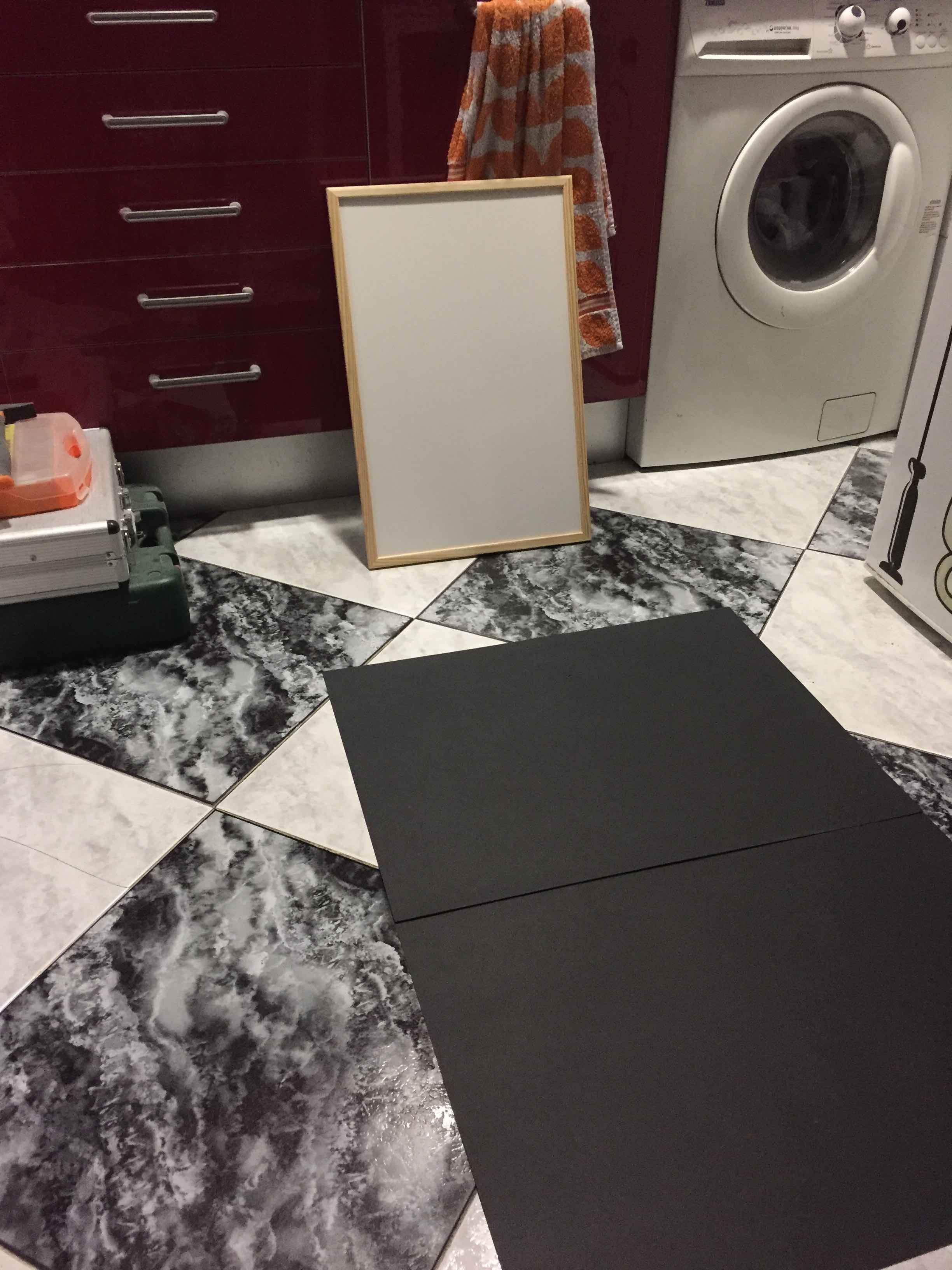}}\\ \vspace{-0.3cm}
  \subfloat[]{\includegraphics[width=0.35\textwidth]{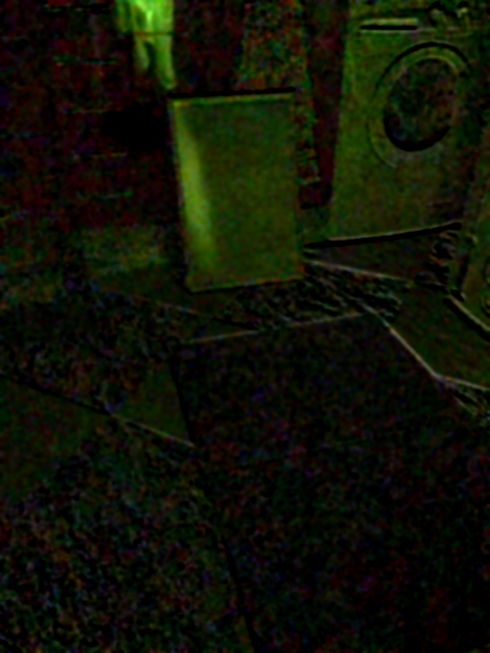}} \hspace{0.1cm}
  \subfloat[]{\includegraphics[width=0.35\textwidth]{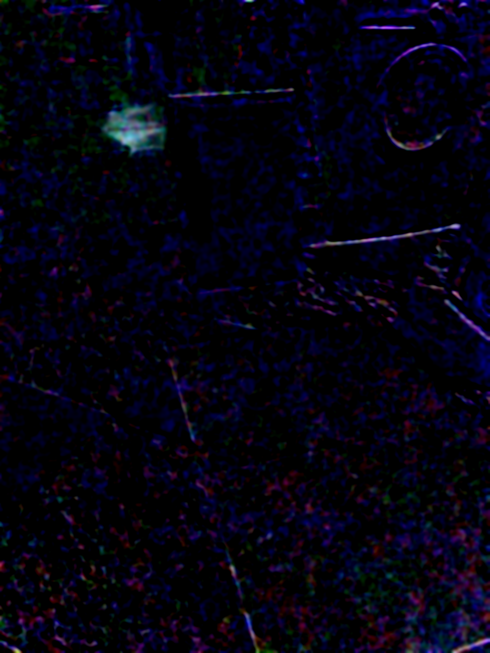}}
  \caption{Identification of a cameraman. (a) Apartment scene with the cameraman's green jacket shown. The jacket was present in the scene when $p$ was acquired and absent from the scene when $p_r$ was acquired. (b) Available scene image $p$. (c) Reference baseline image $p_r$ acquired post-hoc. (d) Difference image $D_{+}$. (e) Difference image $D_{-}$. Images and contents thereof are best seen magnified in an e-copy of our paper.}
\label{fig: ResultsCaseStudy1}
\end{figure}

The scene (Figure \ref{fig: ResultsCaseStudy1}a) consists in an apartment where a cameraman took the picture $p$. For convenience, due to our particular scene configuration, 
we emulated the fact that the jacket was worn by a standing cameraman by leaving that jacket hanging on a coat rack rod of appropriate height near from the camera. This approximation will be dropped in the experiment of Section \ref{Section: VideoApplications}.

The available scene image $p$ does not include the jacket in its FoV (Figure \ref{fig: ResultsCaseStudy1}b). In this image, visual evidence from the jacket is hard to localize for an observer: a reflection of the jacket can only be perceived when zooming and closely inspecting that image in the upper-drawers area. The question is thus whether our strategy can further help locating this visual evidence in $p$ or further retrieve and amplify additional evidence.

Following our strategy, a baseline reference image $p_r$ was first acquired in the same conditions as $p$, except for the fact that no jacket was present in the scene this time (Figure \ref{fig: ResultsCaseStudy1}c). Subsequently, the corresponding difference images $D_{+}, D_{-}$ were generated following the proposed analysis method described in Section \ref{subsect: DiffImageAnalysis}. Our results show that the positive difference image $D_{+}$ (Figure \ref{fig: ResultsCaseStudy1}d) makes the reflections of the green jacket much more straightforward to localize by a human observer on the drawers near the top left of the image. Details such as its shape---as folded around a metallic bar---also correspond to the actual shape of the jacket (Figure \ref{fig: ResultsCaseStudy1}a). In addition, a reflection of the same jacket is also retrieved in $D_{+}$ on the plastic white board located in the middle of the image, unlike in $p$ where this evidence remains invisible even upon close visual inspection. In the negative difference image $D_{-}$ (Figure \ref{fig: ResultsCaseStudy1}e), a small area with significant differences is also observed near from the position of the jacket reflections. These differences may originate from a slight occlusion effect that either the jacket itself or the structure on which it is hanged was creating in the scene.

This experiment shows that DIF can  retrieve and amplify visual evidence and features, such as specific color, shape and cloth information, that can help identify key individuals in a scene. Even in cases where upon close inspection, some visual evidence is already visible in the original picture $p$, it is made much more straightforward to identify and locate, once amplified by our method. 

\section{Video Analysis}
\label{Section: VideoApplications}

DIF and the proposed analysis method are naturally suited to analyze video data. To demonstrate this applicability, 
we conducted an experiment where our iPhone camera was set to a fixed location and orientation, thus mimicking the setup of some surveillance-camera systems. While the apartment inside which the video was acquired and the green jacket are the same as in Section \ref{subsect: SecondCaseStudy}, this experiment involves an intruder who wears this jacket and who may be located at arbitrary positions in the scene, rather than a cameraman who stands behind or near the camera.

\begin{figure}
  \centering
  \subfloat[]{\includegraphics[width=0.7\textwidth]{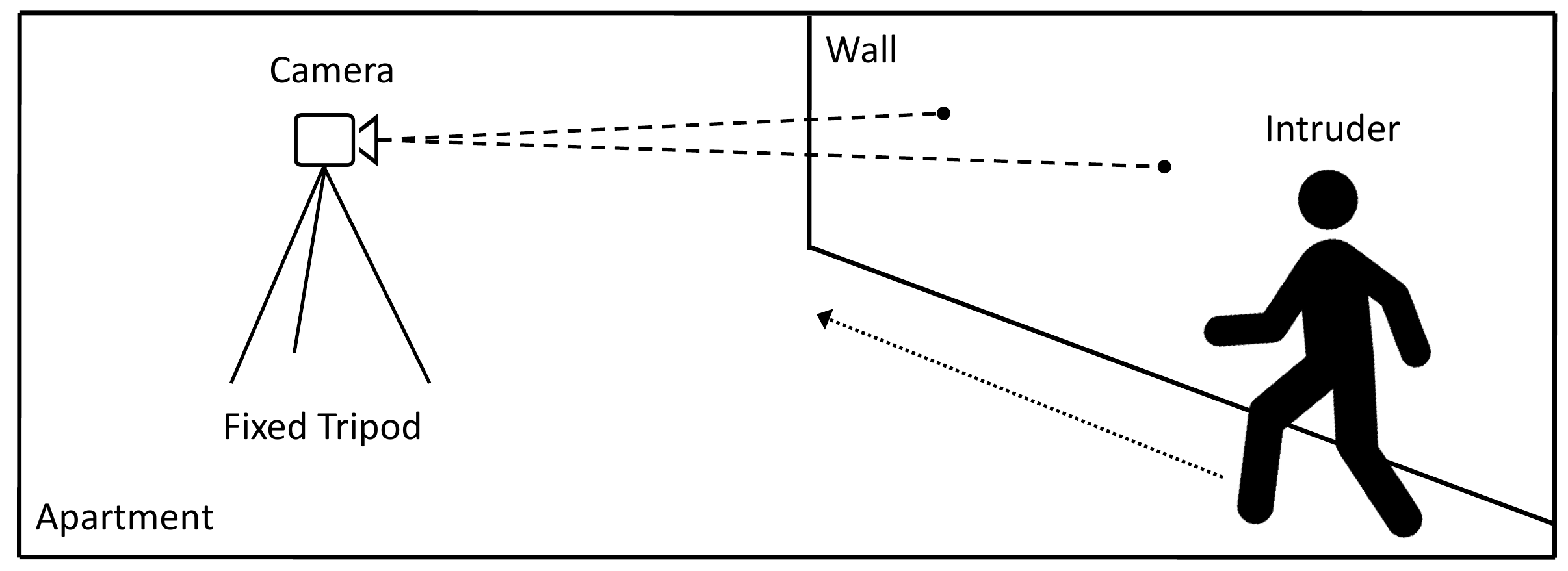}}\\   
  \subfloat[]{\includegraphics[width=\textwidth]{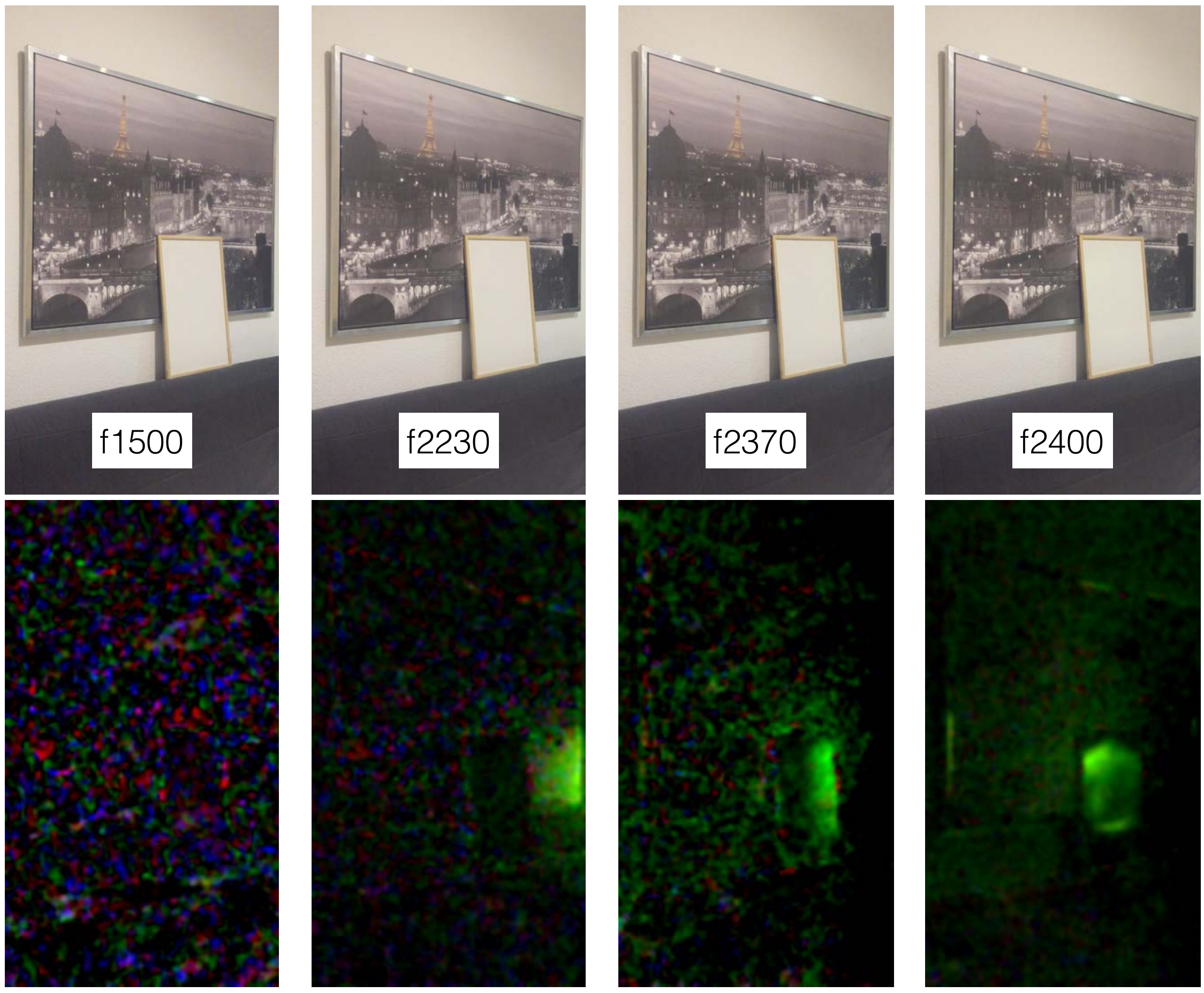}}   
  \caption{Video experiment. (a) Schematics of the scene setting with the iPhone camera pointed towards the wall on its left. The space between the wall and the camera is sufficient for the passage of the intruder in-between. (b) Available scene frames $p^i$ (top row) and corresponding difference scene frames ${D^i_{+}}$ (bottom row). From left to right: (1) the intruder stands outside the apartment with the door closed, (2) the intruder has just entered the apartment, (3) the intruder approaches the plastic white board, and (4) the intruder stands in front of the plastic-white-board surface. All the video frames and contents thereof are best seen magnified in an e-copy of our paper.}
\label{fig: VideoExperiment}
\end{figure}

According to our setup (Figure \ref{fig: VideoExperiment}a), we recorded a raw video of $125$ seconds ($3657$ frames) at $30$ frames per second using the H264 codec. For visualization and analysis, we then exploited the relevant $30$-second chunk of this video (frames $1500$-$2400$) where the intruder first stands outside the apartment with the door closed, then enters the apartment, and finally approaches the camera along the path situated between the camera and the wall. Note that, in this experiment, the reference baseline image $p_r$ was directly extracted from the raw video stream before the frames of interest for convenience. Obviously, doing so implicitly assumes the knowledge of specific video intervals during which no hidden individual or object was present in the scene and which can be used as reference. If no such assumption can be made, $p_r$ can either be acquired post-hoc or from some previous video footage.

Since the frames of interest consists of separate images $p^i$ that can be analyzed with respect to one same reference baseline image $p_r$, our video-analysis problem can be simplified to the single-image case addressed in Section \ref{subsect: DiffImageAnalysis}. In this video setting, however, we slightly adapted our approach to exploit the temporal redundancy between video frames and further reduce noise. First, we computed $p_r$ as the temporal average of Frames $1100$-$1300$ of the raw video, during which the intruder was known to be absent from the scene. Second, for differential image analysis, we used the same spatial-filtering method as in Equation (\ref{eq: Integration}) but used it in conjunction with an additional temporal-filtering operation, using a uniform filtering window of $11$ frames.

Results of this experiment (Figure \ref{fig: VideoExperiment}b) show four selected frames of the video chunk $p^i$ along with their corresponding positive difference images ${D^i_{+}}$. Human observers cannot infer the presence of an intruder in any of these single frames $p^i$; only a careful cross-frame examination reveals subtle temporal color variations. In the positive difference images ${D^i_{+}}$, however, the entrance of the intruder in the apartment creates a clear transition in the corresponding set of frames associated with the amplified reflections that the jacket creates in the scene. Specifically, these amplified reflections gives away green-color hints as visual evidence. In the last difference image shown, one can also recognize the shape of the jacket worn by the intruder.

This experiment demonstrates the capabilities and relevance of our method when dealing with the analysis of video data. The proposed strategy is particularly well suited for videos acquired from fixed camera setups because, in these cases, the reference baseline image can be conveniently acquired with the same device, at any time, and without requiring any adjustment or calibration.

\section{Forgery Detection}
\label{Section: ForgeryDetection}


DIF offers a novel approach to 
forgery detection, \emph{i.e.}, verifying whether an image has been tampered with. 
As shown in Figs.\ref{fig: ResultsCaseStudy2}, \ref{fig: ResultsCaseStudy1} and \ref{fig: VideoExperiment}, we can recover latent visual evidence that is hard or impossible to detect for human observers unless it is subsequently retrieved by our method. From the perspective of forgery detection, the fact that this type of evidence is invisible to the human eye---unlike visible shadows \cite{zhang2009detecting, kee2013exposing} or lighting \cite{johnson2007exposing}---is a crucial advantage. Indeed, such invisible evidence is highly unlikely to have been altered by forgers. The new visual features that our method provides thus constitute a reliable reference to detect potential forgeries.

For instance, let us suppose that the original scene image $p$ also includes the green jacket of Fig. \ref{fig: ResultsCaseStudy1}a in its FoV, but with a color altered by a forger following its acquisition. In this case, similar latent visual evidence as the one revealed in Fig. \ref{fig: ResultsCaseStudy1}d would still be present in the picture. Unlike the jacket itself, this latent evidence would not be tampered with because of its lack of visibility to a human observer. Our method could thus be used to reveal inconsistencies between the color of the forged jacket and the recovered reflections of its unaltered version, thereby revealing the forgery to investigators. 
For forgery detection, one would proceed as in Section \ref{Section: Experiments} except that the scene image $p$ would be first split into two parts: the first part $p'$ similar to Fig. \ref{fig: ResultsCaseStudy1}b, which is analyzed and used to generate the difference image, and the second part $p''$ containing the forged jacket, which is tested against the difference image. In this setting, splitting is necessary because our analysis method is specifically designed to extract and amplify subtle image differences. Without splitting, the forged jacket would become the dominant image difference in Equation (\ref{eq: ContrastNorm}), and more subtle differences would not be sufficiently amplified to become visible. Accordingly, we would first acquire a reference image $p_r'$ reproducing the acquisition parameters of $p'$, and then use our differential-analysis method to generate a difference image $D'_+$ from $p'$ and $p_r'$. Finally, we would compare $D'_+$ with $p''$ to reveal the inconsistencies. 
We note that the exact same method can be used for detecting shape alterations, instead of color modifications, in the image.

The DIF paradigm may also be employed to catch fake video clips generated by artificial intelligence—so-called \emph{deep fakes}—which are increasingly resistant to detection and a growing concern for privacy and security \cite{chesney2018deep}. More specifically, difference images capture additional visual cues created by reflective objects in the scene, the flat-board of Figure \ref{fig: VideoExperiment} being an illustration thereof. Since those subtle impacts are also difficult to detect or invisible to human observers in videos, they are not expected to be reproduced or altered in fake videos. For instance, consider the video footage of an important public figure who enters a room and makes a speech. If this video were a deep fake, the person's face and clothing could have been artificially modified to provide false clues on the person's true identity. In this case, however, potential reflections of the actual person on the walls, floor, or other objects could still be amplified with our method, following a similar splitting strategy as above. This could reveal cues such as the correct clothes color, spot out inconsistencies, and identify the video as a fake. From the perspective of forgery detection, video data also has the advantage of being composed of multiple frames, which implies that difference images can be readily generated from a single footage 
without requiring access to the real scene.


\section{Discussions}
\label{Section: Discussion}



Arguably, the use of DIF may increase the discrepancy between an individual's awareness of the visual evidence he or she may leak through, such as visible shadows, and the strategic/technical capabilities that investigators have at disposal to extract said evidence. Conversely, this refined knowledge could be positively exploited in operations conducted by intelligence experts or agencies. Another relevant aspect of the latent evidence we are considering is that it may also be more immune to tampering, knowing that it can only be revealed through a-posteriori in-depth analysis.

From a technical perspective, one strength of our image-analysis technique originates from its combination between the cross-image comparative process of Equation (\ref{eq: Differentiation}), which is a differentiation process, with the spatial-filtering process of Equation (\ref{eq: Integration}) used for contrast enhancement, which is an integration process. In some sense, the way our proposed technique can reveal a-priori-hidden details through the joint use of these two techniques bears similarities with the so-called Eulerian Video Magnification (EVM) \cite{wu2012eulerian}. The EVM technique, albeit proposed to amplify periodic temporal variations in videos, which differs from our setting, is known to successfully reveal information that cannot be perceived by human observers. For instance, it was applied to recover heart-rate information from imperceptible face-color changes in time \cite{wu2012eulerian}, or even sound from vibrating objects \cite{davis2014visual}. However, the DIF concept and methods that we have proposed are the first for revealing a novel category of latent visual evidence that is not perceivable by human observers in single images or in videos. 

DIF can also be applied to traditional film photography if the photo prints are digitized for subsequent analysis by our method. The process of printing and scanning that traditional photos go through might introduce noise as well as color or brightness alterations. However, to the extent that the process used for the reference image $p_r$ match that used for $p$, these distortions would be similar in both image files and could thus be systematically cancelled via pre-processing.


Several research directions could be followed in further work, which would extend our concept and address limitations of the techniques proposed in this paper. 
For instance, in our current acquisition framework, it is implicitly assumed that both images are acquired from the same position, under the same scene setting and illumination, and with the same camera type and parameters, including aperture, zoom, white balance and focus. However, relaxing this strict requirement would make DIF more practically applicable in some scenarios. This calls for technical refinements where both the device and the parameters of the scene acquisition can vary to some extent for the second image.

Moreover, the algorithmic techniques that we currently apply to extract the latent evidence from image pairs $\{p, p_r\}$ could be substantially improved. For instance, more advanced algorithms could involve better noise-reduction techniques and the use of machine learning, such as artificial neural networks, to identify specific evidence stemming from particular objects in a scene. In the case of videos, temporal redundancy could be better exploited, and difference-image information could also be extracted from arbitrary (\emph{e.g.}, adjacent) frame pairs---as opposed to merely using the same reference image as in Section \ref{Section: VideoApplications}---to provide additional information. Enhanced analysis capabilities would in turn allow to explore the applicability of DIF to more complex scenarios and perhaps recover visual evidence in cases where the objects of interest have less impact on the scene images than in our experiments. 
Finally, our experiments do not yet provide a complete answer to the challenging research question of `who was  behind the camera, given a single available scene image?'. We envisage that our DIF paradigm can be improved to help deduce other physical characteristics of the cameraman such as her body size (width and height) and even clothing materials (what fabric), as our earlier simulation studies suggested \cite{yan2017poster}. These aspects remain interesting and important future research.

Section \ref{Section: ForgeryDetection} suggested that deep fakes could be identified by spotting out inconsistencies between image features that are already visible in the video and details recovered by DIF in the corresponding difference images. In general, two types of scenarios are relevant for deep-fake detection. First, we have an authentic video footage and a modified one, but do not know which is which. In this case, the authentic version could be identified as the one where the visible features are closest to the details generated by DIF. Second, we have only a single video footage that may or may not be a deep fake. In this case, the video could be identified as a deep fake if inconsistencies are detected above the noise level. In both scenarios, difference images could be extracted using a single reference image per video, as in Section \ref{Section: VideoApplications}. Alternatively, comparisons may be made by extracting difference images from adjacent video frames. How to best detect deep fakes with DIF is interesting future work.

\section{Conclusions}
\label{Sect: Conclusions}

This work is the first to introduce the notion and methodology of DIF. Our experiments have demonstrated that DIF can successfully recover a novel category of latent visual evidence in images and video footage, which would otherwise remain difficult to identify by or even invisible to human eyes. 
In principle, DIF is  
applicable to both digital and traditional photography. 

A key insight underlying DIF is that, when a scene is photographed or videotaped, an object that is out of the FoV often deflects light into the scene via reflection and refraction, leaving subtle impacts on the resulting photo or video footage. By introducing an additional baseline reference image, which is similar but not impacted by the object, we can compute image differences, and derive crucial clues for deducing some physical characteristics of the object. In a sense, the object gives away its own presence via an optical side channel, which was entirely ignored in prior work.

While DIF is related to image forensics, they are not the same. Image forensics is a well-established field that largely addresses integrity issues, \emph{i.e.}, whether an image is doctored or forged. The concept of DIF overlaps with image forensics in that the former offers new methods for forgery detection. However, DIF goes well beyond the conventional scope of image forensics, as demonstrated by our experiments 
in Sections \ref{Section: Experiments} and \ref{Section: VideoApplications}. 

DIF connects several well-established areas with each other, such as image forensics, side channels, physical security and information security, via new links. DIF also connects with authorship identification and privacy protection by offering a solution to the rarely studied problem of photographer identification.

Overall, DIF appears to open a new form of digital forensics, with ample opportunities for further explorations. It has practical applications in crime investigation and surveillance, and it also informs intelligence operations. 
%

\bibliographystyle{alpha}
\bibliography{bibliography}

\end{document}